\documentclass[a4paper,11pt]{article}

\usepackage{graphicx}
\usepackage{amsmath}
\usepackage{fancyhdr}
\usepackage{fancyhdr}

\newcommand\Tr{\mbox{Tr}}            
 
\begin{document}

\title{Mathematical estimates for  the attractor dimension in MHD turbulence }
\author{A. Poth\'erat and T. Alboussi\`ere\\
Cambridge University Engineering Department,\\
Trumpington street, Cambridge CB2 1PZ, UK}
\date{3 April, 2003}

\maketitle

\section{Introduction}
The aim of the present work is to derive rigorous estimates for turbulent
MHD flow quantities such as the size and anisotropy of the dissipative 
scales, as well as the transition between 2D and 3D state.
To this end, we calculate an upper bound for the attractor dimension of the 
motion equations, which indicates the number of modes present in the fully 
developed flow. This method has already been used successfully to 
derive such estimates for 2D and 3D hydrodynamic turbulence as a function of the
$\mathcal L_\infty$ norm of the dissipation, as in 
\cite{doering95}. We tackle here the problem of a flow periodic in the 3 
spatial directions (spatial period $2\pi L$), to which a permanent magnetic field is applied. In addition,
the detailed study of the dissipation operator provides more indications about 
the structure of the flow.\\
In section 2, we review the tools of the dynamical system theory as well as the 
results they have led to in the case of 3D hydrodynamic turbulence. Section 3 is 
devoted to the study of the set of modes which minimises the trace of the operator associated
to the total dissipation in MHD turbulence (viscous and Joule). Eventually, the 
estimates for the attractor dimension and dissipative scales in MHD turbulence under
strong magnetic field are derived in section 4 and compared to results obtained from
heuristic considerations.

\section{The Navier-Stokes equation as a dynamical system}
We shall first explain the interest of studying the dynamical system associated
 with the Navier-Stokes equations.
 The quantity we are mostly interested in is the set of functions which 
 "attracts" any initial flow, in the sense of the limit when the time $t$ tends to infinity. Indeed, the dimension of this  so-called
 global attractor is known to be high for turbulent flows, but finite under the assumption that the 
 Navier-Stokes equations do not produce any finite time singularity \cite{doering95}. 
 Physically, this indicates that an established homogeneous turbulent flow 
 includes a finite number of vortices, which therefore cannot be smaller than 
 the ratio of the the volume of the physical domain by the number of modes, 
 precisely given by  the attractor dimension $d_M$. Evaluating an upper bound
 for $d_M$ is thus  a way to derive a lower bound for the size of the 
 dissipative scales. This will be our purpose from now on.\\

 \subsection{Dimension of the attractor associated to the Navier-Stokes equation}
	To calculate the attractor dimension of a dynamical system
(defined by an evolution equation of the kind $\frac{\partial}{\partial t} \mathbf u=\mathbf F(\mathbf u) $), we consider
a solution $\mathbf u$ located on the attractor and an arbitrary number $n$ of
small independent disturbances $\delta \mathbf u_i / i\in \{1..n\}$.
 Note that "small" is relative to the norm defined in the phase space, 
which is a  space of functions in the case of the Navier-Stokes system. The 
subset spanned by these $n$ independent vectors evolves as to be  
located within the attractor at infinite time. Therefore, if $N>d_M$, 
the $n$-dimensional volume of this subset, defined as
\begin{equation}
V_n(t)=|\delta \mathbf u_1 \times..\times \delta \mathbf u_n|
\end{equation}
tends to $0$ when $t$ tends to infinity. This latter property is expressed by
 Constantin and Foias theorem \cite{constantin85_ams}.\\
In the vicinity of the attractor, the evolution operator can be linearised as 
$\mathbf F(\mathbf u)=\mathbf{Au}+O(\|\delta\mathbf u\|^2) $ so 
that $V_n(t)$ varies exponentially in time:
\begin{equation}
 V_n(t)=V_n(0)\exp(t\langle\Tr(\mathbf{A_n})\rangle)
 \label{volume}
\end{equation}
The subscript $n$ stands for projection of operator into $n$-dimensional subsets
 of the phase space.
 If $\Tr(A_n)$ is positive for at least one choice of $n$ disturbances, then $n$
 is an upper bound for the attractor's dimension because at least one
 $n$-volume would expand (see (\ref{volume})).
 We shall therefore look for the maximum trace of the 
 evolution operator associated to the Navier-Stokes equations for any arbitrary integer $n$.\\


If $\sigma$ is the electrical
conductivity, $\rho$ is the density, $\nu$ is the kinematic viscosity, the motion equations for
velocity $\mathbf u$, pressure $p$ electric current density $\mathbf j$ can be written:
\begin{eqnarray}
(\partial_t + \mathbf u .\nabla)\mathbf u + \frac{1}{\rho}\nabla p
&=& \nu (\nabla^2 \mathbf  u + \frac{\sigma}{\rho \nu} \mathbf j \times \mathbf B) + \mathbf f
\label{eq:ns}\\
\nabla . \mathbf u &=& 0
\end{eqnarray}
where $\mathbf f$ represents some forcing independent of the velocity field.
The set of Maxwell equations as well as electric current conservation and the Ohm's law
are normally required to close the system. However, we assume here that the magnetic field is
not disturbed by the flow. In other words, the magnetic diffusion is supposed to take place
instantaneously at the time scale of the flow ("low magnetic Reynolds number" approximation).\\
In the literature, the inertial terms are often written as a bilinear operator
$\mathcal B(\mathbf u,\delta \mathbf u)$, and the dissipation, as a linear
 operator that we call $\mathcal D_{Ha}$.
 One can guess from this equation, that the evolution of small volume of the phase space
 generated by a set of $n$  disturbances
 (as defined in section previously) results from the competition between inertial terms which
 tend to expand the volume by vortex stretching and dissipative terms
 which tends to damp the disturbances, and hence reduce the volume.
\subsection{The case of hydrodynamic turbulence ($\mathbf B=0$)}
The case without magnetic field ($\mathbf B=0$) has been investigated in 2 and 3 dimensions. In 2D,
\cite{doering95} found an upper bound for the attractor dimension which matches well
the results obtained by Kolmogorov-like arguments.
To this day, no rigorous estimate for the attractor dimension of the 3D problem precisely matches Kolmogorov's
prediction for the number of degrees of freedom.
One of the main reasons is that unlike in 2D, it has not
yet been proved that the velocity gradients remain finite at finite time, which lets the door open
to possible singularities. However, one can work under the assumption that the flow remains regular
at finite time and define the maximum local energy dissipation rate as:
\begin{equation}
\epsilon_\infty = \nu \langle \sup_{\mathbf u} \sup_{\mathbf r} \|\nabla \mathbf u (\mathbf r,t)\|^2 \rangle_t
\label{eq:eps}
\end{equation}
Here, $\sup_{\mathbf u}$ stands for the upper bound over the set of solutions $\mathbf u$ in the phase space,
whereas $\sup_{\mathbf r}$ stands for the upper bound over the physical domain.
Under this strong assumption, and using a typical large scale $L$,
which can be extracted from the eigenvalue
of the laplacian of smallest module $-\lambda_1$, such that $L=\lambda_1^{-1/2}$
an upper bound for the trace of the
operator $\mathcal B(.,\mathbf u)$ on any $n$-dimensional subspace of the phase
space is presented in \cite{constantin85_jfm}:
\begin{equation}
|\Tr(\mathcal B(.,\mathbf u))|< \frac{\nu}{L^2} \left(\frac{\epsilon_\infty L^4}{\nu^3}\right)^{1/2}
\label{eq:trb}
\end{equation}
Also, studying the sequence of eigenvalues of the dissipation operator (which reduces to a Laplacian
in the absence of magnetic field) on a finite physical domain with appropriate boundary conditions,
gives access to the trace of the dissipation operator (see for instance \cite{doering95})
and provides an upper bound for the trace of the total evolution operator, on any $n$-dimensional
subspace of the phase space:
\begin{equation}
\Tr((\mathcal B(.,\mathbf u)+ \nu \nabla^2)\mathbf P_n) \leq \nu \lambda_1 n (\left(\frac{\epsilon_\infty L^4}{\nu^3}\right)^{1/2}-c n^\frac{2}{3})
\label{eq:trace_constantin}
\end{equation}
where $c$ is a real constant of the order of unity.One can be sure that 
when $n$ is such that the \textit{r.h.s.} of (\ref{eq:trace_constantin}) is
negative, all $n$-volumes shrink, hence $n>d_{3D}$ where $d_{3D}$ is the attractor's dimension
(this is Constantin and Foias theorem \cite{constantin85_ams}). It then comes from  (\ref{eq:trace_constantin}) that:
\begin{equation}
d_{3D} \leq c_3 \left(\frac{\epsilon_\infty L^4}{\nu^3}\right)^{3/4}
\label{eq:dim3D}
\end{equation}
The bound (\ref{eq:dim3D}) apparently matches the Kolmogorov estimate of
$\left(\frac{\epsilon L^4}{\nu^3}\right)$. Unfortunately, the maximum local dissipation
defined in (\ref{eq:eps})
could be much higher than the average dissipation rate used in the Kolmogorov theory \cite{k41}. Note
that a more recent attempt to find an upper bound for $d_{3D}$ \cite{gibbon97} using the average dissipation
$\bar\epsilon$  has led to $d_{3D}<\left(\frac{\bar\epsilon L^4}{\nu^3}\right)^{\frac{24}{5}}$. We will however still
 use (\ref{eq:trb}) throughout the rest of this work as this bound turns out to be relevant when a strong
   magnetic field is applied to the flow (see section \ref{sec:bounds}). Note also that the discrepancy
     between analytical estimates and heuristic results is due to the difficulty in getting
     estimates for the norms of the velocity gradients, as well as to the fact that the bound given here
     does not rely on the existence of a power-law spectrum, which makes it also valid for flows with a low Reynolds number,
     unlike the K41 theory \cite{k41}.\\
     
In order to derive an estimate for the attractor dimension in the MHD case, our
 main task now consists in finding the minimum of the trace of the dissipation
 operator on all $n$-dimensional subspace, for arbitrary values of $n$.

\section{Properties of the modes minimising the dissipation}
\begin{figure}
\begin{center}
\includegraphics[scale=0.65]{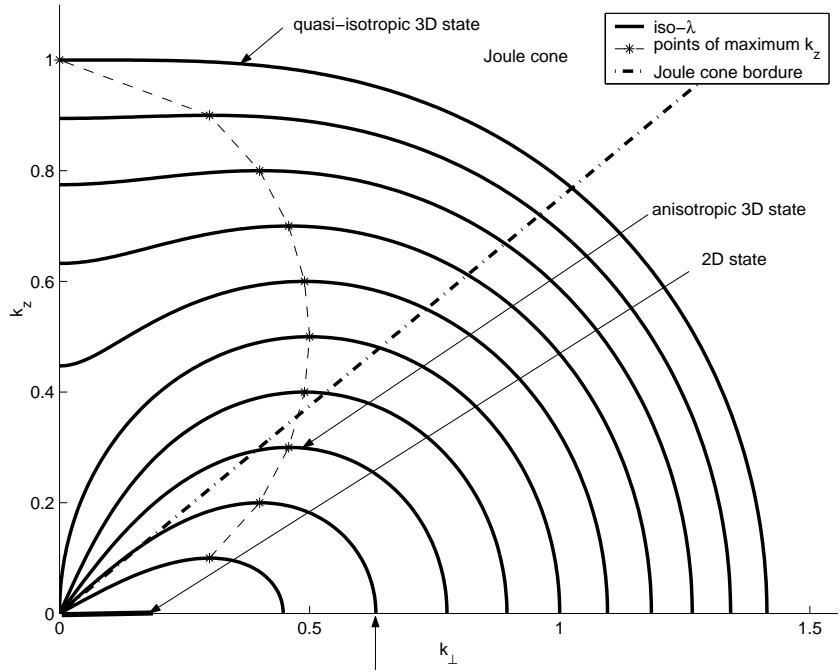}
\caption{Iso-$\lambda$ curves in the plane ($k_\perp,k_z$). One can see the three major types of
mode distribution: the 2d state corresponds to a set of modes located on the $k_\perp$ axis, the
strongly anisotropic 3d state exhibits the Joule cone-like shape (the bordure of the Joule cone
has been plotted in the case where all the modes are inside the curve designated by the vertical arrow)
and the quasi-isotropic state is reached when the modes are enclosed inside curves located the furthest
away from the origin. Axis units are arbitrary.
}
\label{graphs}
\end{center}
\end{figure}

We shall now look for the set of $n$ modes that achieve the minimum 
dissipation for any value of $n$ and exhibit a few important properties of 
these modes. The dissipation operator is compact and self-adjoint, so 
its trace expresses as the sum of its eigenvalues. The next step is now to 
solve the eigenvalue problem of the dissipation operator and to sort the 
eigenvalues in ascending order. The sum of the $n$ first actually achieves the 
minimum of the trace over all $n$-dimensional subset of the phase space. Using
non dimensional dissipation operator and wavenumbers (normalised respectively 
by $\frac{\nu}{L^2}$ and $L^{-1}$), the 
three spatial component of the eigenvector appear to be of the form:
\begin{equation}
U(x,y,z)=\sin(k_x x)\sin(k_y y)\sin(k_z z) / (k_x,k_y,k_z)\in 
N^3 (k_x,k_y,k_z) \neq (0,0,0)
\end{equation}
The eigenvalue associated to the mode $(k_x,k_y,k_z)$ expresses its dissipation
rate and writes:
\begin{equation}
\lambda(k_x,k_y,k_z)=-(k_x^2+k_y^2+k_z^2)-Ha^2\frac{k_z^2}{k_x^2+k_y^2+k_z^2}
\label{dissipation}
\end{equation}
where the square of the Hartmann number $Ha^2=L^2B^2\frac{\sigma}{\rho\nu}$
represents the ratio of Joule to viscous dissipation.
The function $\lambda(k_x,k_y,k_z)$ is convex so that if $\lambda_m$ is the 
largest eigenvalue (corresponding to the $n^{th}$ mode), all modes associated to
smaller eigenvalues are located inside the area delimited by the curve 
$\lambda(k_\perp,k_z)=\lambda_m$ in the $(k_\perp,k_z)$ plane, where 
$k_\perp=\sqrt{k_x^2+k_y^2}$, as shown on figure \ref{graphs} . The 
knowledge of the iso-$\lambda_m$ curve also provides the maximum values of 
the modes in the direction of the magnetic field $k_{z_m}$ and in the 
orthogonal 
direction $k_{\perp_m}$, the ratio of which is an indication of the 
anisotropy of the small scales.\\
These features can be used to calculate the $n$ first modes and 
the associated trace of the dissipation as a function of $n$ and $Ha$. This 
is done in the general case, using an iterative algorithm implemented on a 
computer.\\
The shape of the iso-$\lambda_m$ is determined by the ratio $\frac{n}{Ha^2}$
(see figure \ref{graphs})
. Intuitively, it indicates the relative importance of forcing versus 
dissipation (as a higher inertia tends to generate more modes, and hence, 
increase the dimension of the attractor). We notice that the smaller this 
number, the more modes are concentrated outside of the a cone of axis (Oz). 
This behaviour has been pointed  out both experimentally \cite{alemany79} and 
theoretically \cite{sm82} for real flows, for which a strong magnetic field is
known to result in turbulent modes being confined outside the Joule cone. 
For dominating 
electro-magnetic effects, the Joule cone extends to the whole space except from 
the horizontal plane $(kx,ky)$ : the flow becomes two-dimensional. This also 
occurs in the eigenvalue problem where two-dimensional modes appear to be the 
less dissipative ones. This allows us to find out whether the set of $n$ 
eigenmodes is purely two-dimensional (\textit{i.e} when all the modes satisfy
$k_z=0$).\\ 
In the case of a distribution of a high number of 3D modes ($n>>1$) located 
outside of the Joule cone ($Ha<-\lambda_m<Ha^2$). An analytical expression for the trace of the 
dissipation, as well as the Joule cone angle $\theta_m$ can be found, by replacing the sum 
over the $n$ eigenvalues by an integral \cite{pa03}:
\begin{eqnarray}
\Tr(\mathcal D_{Ha}\mathbf P_n)=\frac{2\sqrt{2}}{3\pi}n^{3/2}Ha^{1/2}
\label{trace_d}\\
\sin \theta_m=\sqrt{\frac{2}{\pi}}n^{1/4}Ha^{-3/4}
\label{eq:theta}
\end{eqnarray}
The geometrical shape of the cardioid yields the 
maximum wavenumbers in the z-direction and in the orthogonal direction:
%
\begin{eqnarray}
k_{\perp_m}&=& \sqrt{-\lambda_m}=\frac{2^{1/4}}{\pi^{1/2}}n^{1/4}Ha^{1/4}
\label{eq:kpmax0}\\
k_{z_m}&=&-\frac{\lambda_m}{2Ha}=\frac{1}{\pi\sqrt{2}}n^{1/2}Ha^{-1/2}
\label{eq:kzmax0}
\end{eqnarray}
and the set of minimal modes  is two dimensional if and only if $n<2\pi^2 Ha$.
The properties of the eigenmodes of the dissipation  operator and those of
the real flow exhibit some striking similarities. We shall exhibit more of them 
using the full result on the estimate for the attractor dimension.

\section{Bounds on turbulent MHD flow quantities}
\label{sec:bounds}
\subsection{Analytical estimates}
\begin{figure}
\begin{center}
\includegraphics[scale=0.65]{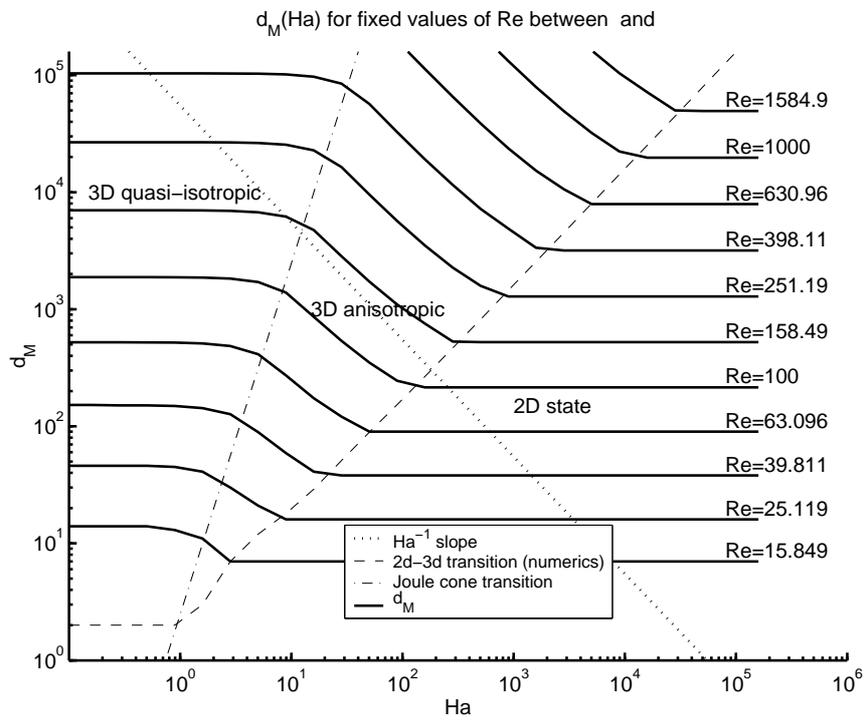}
\caption{Attractor dimension as a function of $Ha$.
dotted: strong field approximation. solid: numerics.}
\label{dim_attract}
\end{center}
\end{figure}

We shall now derive an estimate for the attractor dimension of the Navier-Stokes
 equation on a periodical domain. To this end, we add (\ref{eq:trb}) 
 to the result of the numerical calculation of the trace
 in order to get an upper bound  for the 
 expansion rate of the Volume of any $n$-dimensional subset located in the 
 vicinity of the attractor. We recall that the attractor dimension is the 
 smallest value of the integer $n$ for which this expansion rate is negative.
 The results are plotted on figure \ref{dim_attract} in the general case. In the
 case $n>>1$ and $Ha<-\lambda_m<Ha^2$, abound for the trace of the evolution operator 
 can be expressed analytically by summing (\ref{eq:trb}) and (\ref{trace_d}) so that so that using equations (\ref{eq:kpmax0}-\ref{eq:kzmax0}) 
 we get an analytical upper bound for the attractor dimension, as well as  upper  bounds for the maximum wavenumbers:
\begin{eqnarray}
d_M \leq \frac{9\pi^2}{32}\frac{Re^4}{Ha} 
\label{eq:dimat1}\\
k_{\perp_m}\leq {\frac{\sqrt{3}}{2}} Re
\label{eq:kpmd1}\\
k_{z_m} \leq \frac{3}{8}\frac{Re^2}{Ha}
\label{eq:kzmd1}
\end{eqnarray}
%
%
\subsection{Heuristics on MHD turbulence of Kolmogorov type under strong field}
Now, it is  worth
underlining again that estimates (\ref{eq:dimat1},\ref{eq:kpmd1},\ref{eq:kzmd1})  are exact
results, and come exclusively from the
mathematical properties of the Navier-Stokes equations, without the involvement of any physical
approximation. There is therefore considerable interest in comparing  them with orders of
magnitude obtained from heuristic considerations. Let us recall how the smallest scales can
be obtained in a more intuitive manner: in a 3D periodic flow where Joule dissipation is stronger than
viscosity except at small scales ($Ha>>1$), it is usual to  consider that a vortex 
in the inertial range, of typical velocity $U_v$ and scales $k_\perp$ and $k_z$,
results from a balance between inertial and Lorentz forces, which implies:
\begin{equation}
\frac{k_z}{k_\perp} \sim \left(\frac{\sigma B^2 L}{\rho k_\perp U_v}\right)^{-1/2}
\label{eq:anis}
\end{equation}
Moreover, one usually assumes that anisotropy remains the same at all scales \cite{alemany79},
over the inertial range. Under this assumption, (\ref{eq:anis}) implies  $U_v(k_\perp)=U_0k_\perp^{-1}$,
where $U_0$ stands for a typical large scale velocity.
This is usually expressed in terms of the energy spectrum as:
\begin{equation}
E(k_\perp)\sim k_\perp^{-1} U_v^2(k_\perp) \sim U_0^2k_{\perp}^{-3}
\label{eq:k-3}
\end{equation}
As mentioned in introduction the $k^{-3}$ spectrum is a strong feature of this type
of MHD turbulence. Using the dissipation defined as
$\epsilon \sim \nu k_\perp^2 U(k_\perp)^2 \sim \nu U_0^2$, the large scale velocity 
expresses as
$U_0 \sim \sqrt{\frac{\epsilon}{\nu}}$, and (\ref{eq:anis}) can then be written:
\begin{equation}
\frac{k_z}{k_\perp} \sim \left( \frac{\epsilon L^4}{\nu^3}\right)^{1/4} \frac{1}{Ha}=N^{-1/2}
\label{eq:anis_h}
\end{equation}
where the ratio $N$ is the interaction parameter, which represents the ratio
between Lorentz forces and inertia. Eventually, the small scales  are
heuristically defined as the smallest possible structures of the inertial range which are not
destroyed by viscosity, which means that they result from a balance between inertia and viscosity.
This yields:
\begin{equation}
\frac{k_{z_m}}{k_{\perp_m}^2} \sim Ha^{-1}.
\label{eq:sscales}
\end{equation}
Now combining (\ref{eq:anis_h}) and (\ref{eq:sscales}) yields:
\begin{eqnarray}
k_{\perp_{max}} \sim  \left( \frac{\epsilon L^4}{\nu^3}\right)^{1/4} \\
k_{z_{max}} \sim  \left( \frac{\epsilon L^4}{\nu^3}\right)^{1/2} \frac{1}{Ha}
\label{eq:kzmp1_h}
\end{eqnarray}
from which the number of degrees of freedom of the flow can be estimated by counting the number of vortices in
the of size $L/k_\perp\times L/k_\perp \times  L/k_z$ in a $L\times L\times L\times$ box:
\begin{equation}
N_f \sim k_\perp^2k_z \sim \frac{1}{Ha} \left( \frac{\epsilon L^4}{\nu^3}\right) \sim d_M
\label{eq:nf_h}
\end{equation}
This suggests that our estimate for $d_M$ is sharp and yields the right order of magnitude for the
small scales. Note also that  (\ref{eq:theta}) and (\ref{eq:dimat1})
yield $\sin\theta_m =\frac{\sqrt{3}}{\sqrt{2}}N^{-1/2}$ which matches the 
heuristic prediction of \cite{sm82} for the Joule cone angle.

\section{Conclusion}
Though they are not solution of the motion equations, the eigenmodes of the dissipation operator exhibit some strong similarities with what is known from the
real flow. As these properties are derived under the only assumption that the 
solutions of the Navier-Stoles equations are regular, this gives some strong 
support to the assumptions on which former heuristic results rely. However, the
estimates obtained might be improved if the estimate for the inertial terms
 is improved. 
 Indeed, the dissipation defined in (\ref{eq:eps}) is generally higher than  the average
 dissipation used to derive the heuristic value of the $d_M$ so the mathematical
 estimate found for $d_M$ is somewhat too high compared to the heuristics. These 
 rigorous estimates are however very encouraging as they feature the same dependence 
 on the Hartmann number as the heuristic results. This confirms that the set of 
 minimal modes of the dissipation operator do render well the MHD properties of the 
 actual flow.\\
 lastly, it is worth mentioning that  more physical behaviour such as boundary 
 layer velocity 
 profiles could be recovered by performing some similar study with classical
 wall boundary conditions on the  planes orthogonal to the magnetic field.\\
 This work has been supported by the Leverhulme Trust (grant F/09 452/A).


\end{document}